\documentclass[sigconf]{acmart}
\acmDOI{XXXXXXX.XXXXXXX}
\acmConference[Conference acronym 'XX]{Make sure to enter the correct
  conference title from your rights confirmation email}{June 03--05,
  2018}{Woodstock, NY}
\acmISBN{978-1-4503-XXXX-X/2018/06}
\usepackage{booktabs}
\usepackage[table]{xcolor}

\definecolor{benchmarkrow}{gray}{0.88}
\usepackage{colortbl}
\usepackage{multirow}
\usepackage{tikz}
\usepackage[edges]{forest}

\usepackage{booktabs}
\usepackage{array}
\usepackage[table]{xcolor}

\newcolumntype{C}[1]{>{\centering\arraybackslash}m{#1}}
\newcolumntype{L}[1]{>{\raggedright\arraybackslash}m{#1}}

\begin{document}

\title{Agentic Quantitative Trading: A Survey of Workflows, Systems, and Evaluation}

\author{Fengrui Hua}
\authornote{Equal contribution.}
\email{fhua430@connect.hkust-gz.edu.cn}
\affiliation{%
  \institution{The Hong Kong University of Science and Technology (Guangzhou)}
  \city{Guangzhou}
  \country{China}
}

\author{Hengyi Yang}
\authornotemark[1]
\email{hengyiyangyhy@gmail.com}
\affiliation{%
  \institution{HSBC Business School, Peking University}
  \city{Shenzhen}
  \country{China}
}

\author{Xinlei Hao}
\authornotemark[1]
\email{xinleihao@stu.pku.edu.cn}
\affiliation{%
  \institution{HSBC Business School, Peking University}
  \city{Shenzhen}
  \country{China}
}

\author{Haohan Zhang}
\email{hzhang760@connect.hkust-gz.edu.cn}
\affiliation{%
  \institution{The Hong Kong University of Science and Technology (Guangzhou)}
  \city{Guangzhou}
  \country{China}
}

\author{Bokai Cao}
\email{mabkcao@connect.hkust-gz.edu.cn}
\affiliation{%
  \institution{The Hong Kong University of Science and Technology (Guangzhou)}
  \city{Guangzhou}
  \country{China}
}

\author{Yiyan Qi}
\email{qiyiyan@idea.edu.cn}
\affiliation{%
  \institution{IDEA Research, International Digital Economy Academy}
  \city{Shenzhen}
  \country{China}
}

\author{Jia Li}
\email{jialee@connect.hkust-gz.edu.cn}
\authornote{Corresponding authors.}
\affiliation{%
  \institution{The Hong Kong University of Science and Technology (Guangzhou)}
  \city{Guangzhou}
  \country{China}
}

\author{Jian Guo}
\authornotemark[2]
\email{guojian@idea.edu.cn}
\affiliation{%
  \institution{IDEA Research, International Digital Economy Academy}
  \city{Shenzhen}
  \country{China}
}

\renewcommand{\shortauthors}{Hua et al.}

\begin{abstract} 
Quantitative trading is moving from isolated predictive models toward agentic workflows that combine reasoning, tool use, memory, and feedback. This survey reviews agentic quantitative trading across five stages: factor mining, signal discovery, portfolio construction, order execution, and risk management. We further examine agentic quant trading systems through architecture, coordination, and adaptation, while comparing benchmarks across strategy construction, offline trading, live market evaluation, and reliability assessment. Our review finds that current systems remain concentrated on signal discovery, while complete integration with portfolio construction, execution, and risk control is still uncommon. Multi-agent systems also rely heavily on aggregation despite increasingly diverse workflow structures. Benchmark evidence further shows that strong model or forecasting capability does not reliably translate into trading performance under live market conditions and reliability controls. We conclude with future directions for more complete trading workflows, stronger coordination, and evaluation matched to the capability being assessed.
\end{abstract}


\maketitle

\section{Introduction}  
Quantitative trading refers to the use of mathematical models, statistical analysis, and machine learning algorithms to analyze market data, identify trading signals, and make investment decisions. Over the past decades, it has evolved from human-designed factors and statistical models to deep learning pipelines \citep{4.0}, driven by the increasing scale, complexity, and nonlinearity of financial data. More recently, the need to integrate heterogeneous market information and support complex financial decisions has motivated systems driven by LLMs~\citep{fingpt,golden} and financial foundation models~\citep{kronos,fincast}. However, quantitative trading is not merely a prediction problem. It is a long-horizon workflow involving factor mining, signal validation, portfolio construction, order execution, risk monitoring, and continual adaptation under non-stationary markets.  

In this survey, we use the term \textit{agentic} to refer to LLM-based systems that go beyond one-shot text generation. An agentic system typically decomposes a complex objective into intermediate subtasks, interacts with external tools, maintains task-relevant state, and updates its behavior through feedback, reflection, or iterative planning \cite{abou2025agentic}. These capabilities have made LLM agents increasingly relevant across a wide range of domains, including software engineering \cite{otoum2026methods}, materials science \cite{zhang2026towards}, and financial services \cite{okpala2025agentic}.

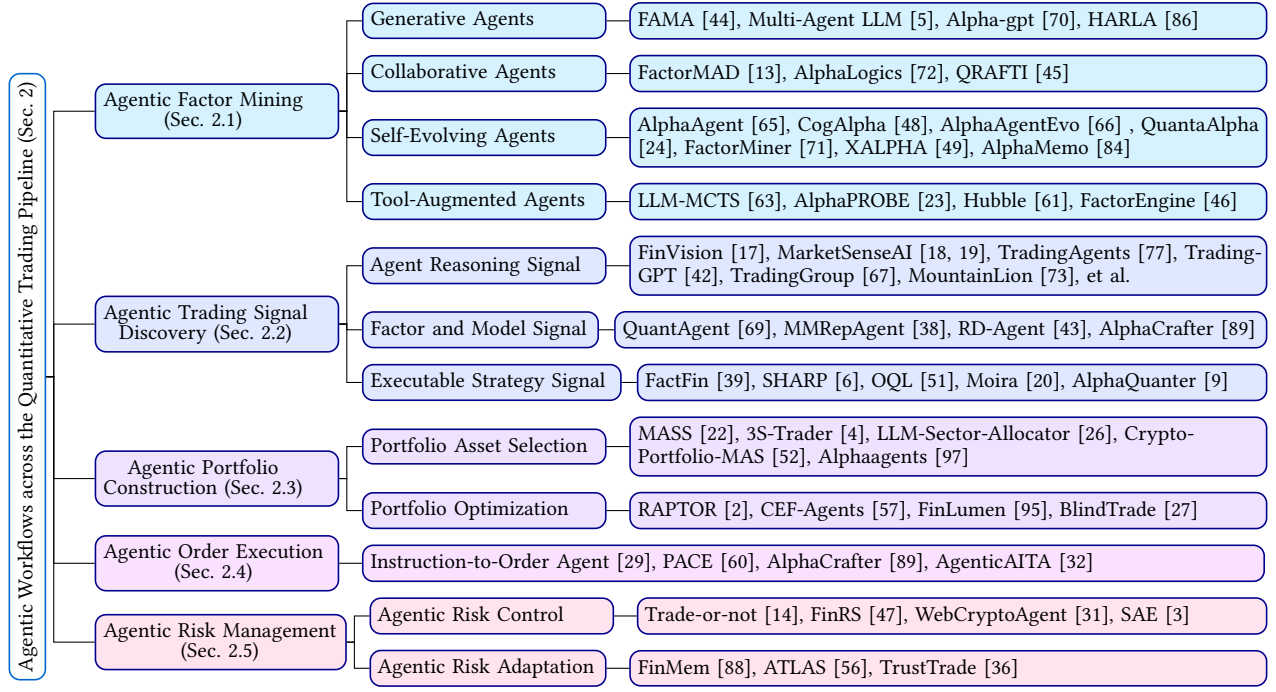
\begin{figure*}[t!]
\centering
\vspace{5mm}
\definecolor{myblue}{RGB}{0,102,204}
\definecolor{blueborder}{RGB}{0,0,139}

\definecolor{rootblue}{RGB}{210,229,255}

\definecolor{gradone}{RGB}{214,242,255}
\definecolor{gradtwo}{RGB}{224,232,255}
\definecolor{gradthree}{RGB}{238,226,255}
\definecolor{gradfour}{RGB}{250,225,255}
\definecolor{gradfive}{RGB}{255,228,240}

\begin{forest}
for tree={   
font=\fontsize{8}{5}\selectfont,
draw=myblue, semithick, rounded corners,
minimum height=1.ex,
minimum width=3em,
anchor=west,
grow=east,
reversed,
forked edge,
s sep=2mm,
fork sep=1mm,
}
[{Agentic Workflows across the Quantitative Trading Pipeline (Sec.~\ref{sec:2})}, rotate=90, anchor=center,l sep=1mm,
    [{Agentic Factor Mining\\(Sec. \ref{sec:2.1})}, 
    text width=3cm, align=center, l sep=3mm, fill=gradone, draw=blueborder
        [{Generative Agents}, text width=3cm, align=center, l sep=3mm, fill=gradone, draw=blueborder
            [{FAMA \citep{FAMA}, Multi-Agent LLM \citep{Multi-Agent-LLM}, Alpha-gpt \citep{Alpha-gpt}, HARLA \citep{HARLA}}, text width=8.2cm, fill=gradone, draw=blueborder]
        ]
        [Collaborative Agents, text width=3cm, align=center, l sep=3mm, fill=gradone, draw=blueborder
            [{FactorMAD \citep{FactorMAD}, AlphaLogics \citep{Alphalogics}, QRAFTI \citep{QRAFTI}}, text width=8.2cm, fill=gradone, draw=blueborder]
        ]
        [Self-Evolving Agents, text width=3cm, align=center, l sep=3mm, fill=gradone, draw=blueborder
            [{AlphaAgent \citep{AlphaAgent}, CogAlpha \citep{CogAlpha}, AlphaAgentEvo \citep{AlphaAgentEvo} , QuantaAlpha \citep{QuantAlpha}, FactorMiner \citep{FactorMiner}, XALPHA \citep{XALPHA}, AlphaMemo \citep{AlphaMemo}}, text width=8.2cm, fill=gradone, draw=blueborder]
        ]
        [Tool-Augmented Agents, text width=3cm, align=center, l sep=3mm, fill=gradone, draw=blueborder
            [{LLM-MCTS \citep{LLM-MCTS}, AlphaPROBE \citep{AlphaPROBE}, Hubble \citep{Hubble}, FactorEngine \citep{FactorEngine}}, text width=8.2cm, fill=gradone, draw=blueborder]
        ]
    ]
    [Agentic Trading Signal\\Discovery (Sec. \ref{sec:2.2}), text width=3cm, align=center, l sep=3mm, fill=gradtwo, draw=blueborder
        [Agent Reasoning Signal, text width=3cm, l sep=3mm, fill=gradtwo, draw=blueborder
            [{FinVision \cite{FinVision}, MarketSenseAI \citep{MarketSenseAI, Signal-or-Noise}, TradingAgents \cite{Tradingagents}, TradingGPT \cite{Tradinggpt}, TradingGroup \cite{TradingGroup}, MountainLion \cite{MountainLion}, et al.}, text width=8.2cm, fill=gradtwo, draw=blueborder]
        ]
        [{Factor and Model Signal}, text width=2.9cm, l sep=2mm, fill=gradtwo, draw=blueborder
            [{QuantAgent \cite{Quantagent}, MMRepAgent \cite{MMRepAgent}, RD-Agent \cite{RD-Agent-Q}, AlphaCrafter \cite{AlphaCrafter}}, text width=8.4cm, fill=gradtwo, draw=blueborder]
        ]
        [Executable Strategy Signal, text width=3.2cm, l sep=2mm, fill=gradtwo, draw=blueborder
            [{FactFin \cite{FactFin}, SHARP \cite{SHARP}, OQL \cite{OQL}, Moira \cite{Moira}, AlphaQuanter \cite{alphaquanter}}, text width=8.1cm, fill=gradtwo, draw=blueborder]
        ]
    ]
    [ Agentic Portfolio\\Construction (Sec. \ref{sec:2.3}), text width=3cm, align=center, l sep=3mm, fill=gradthree, draw=blueborder
        [Portfolio Asset Selection, text width=3cm, align=center, l sep=3mm, fill=gradthree, draw=blueborder
            [{MASS \citep{MASS}, 3S-Trader \citep{3S-Trader}, LLM-Sector-Allocator \citep{LLM-Sector-Allocator}, Crypto-Portfolio-MAS \citep{Crypto-Portfolio-MAS}, Alphaagents \citep{Alphaagents}}, text width=8.2cm, fill=gradthree, draw=blueborder]
        ]
        [Portfolio Optimization, text width=3cm, align=center, l sep=3mm, fill=gradthree, draw=blueborder
            [{RAPTOR \citep{RAPTOR}, CEF-Agents \citep{CEF-Agents}, FinLumen \citep{FinLumen}, BlindTrade \citep{BlindTrade}}, text width=8.2cm, fill=gradthree, draw=blueborder]
        ]
    ]
    [Agentic Order Execution\\(Sec. \ref{sec:2.4}), text width=3cm, align=center, l sep=3mm, fill=gradfour, draw=blueborder
        [{Instruction-to-Order Agent \citep{Instruction-to-Order}, PACE~\citep{PACE}, AlphaCrafter \citep{AlphaCrafter}, AgenticAITA \citep{AgenticAITA}}, text width=11.7cm, fill=gradfour, draw=blueborder]
    ]
    [Agentic Risk Management\\(Sec. \ref{sec:2.5}), text width=3.1cm, align=center, l sep=3mm, fill=gradfive, draw=blueborder
        [Agentic Risk Control,
            text width=3cm,
            l sep=3mm,
            fill=gradfive,
            draw=blueborder
            [{Trade-or-not \citep{Trade-or-not},
              FinRS \citep{FinRS},
              WebCryptoAgent \citep{WebCryptoAgent},
              SAE \citep{SAE}},
              text width=8.1cm,
              fill=gradfive,
              draw=blueborder]
        ]
        [Agentic Risk Adaptation,
            text width=2.9cm,
            l sep=3mm,
            fill=gradfive,
            draw=blueborder
            [{FinMem \citep{Finmem},
              ATLAS \citep{ATLAS},
              TrustTrade \citep{TrustTrade}},
              text width=8.2cm,
              fill=gradfive,
              draw=blueborder]
        ]
    ]
]
\end{forest}
\caption{Taxonomy of agentic workflows across the quantitative trading pipeline.}
\label{fig:tree_full}
\end{figure*}
\textbf{Why Agentic Quantitative Trading?} Agentic quantitative trading emerges from this workflow-level demand. Unlike isolated predictive models, LLM-based agentic systems can decompose complex trading tasks, coordinate specialized agents, and invoke external tools such as databases and backtesting engines \citep{li2025orchestration}. They can also retain memory of past decisions, reflect on failures, and adapt future behavior according to market regimes and performance feedback \citep{abou2025agentic}. This shifts the central question from whether a model can forecast returns to how an agentic system can automate, revise, and control decisions across the trading process.

\textbf{Comparison with Existing Surveys.}
Existing surveys provide important foundations but adopt different units of analysis. The ICAIF 2025 survey by \citet{saha2025large} reviews LLM agents across investment management and organizes the literature around broad use cases and agent designs. Xia et al.~\citep{xia2026agentic} examine trading agents as decision pipelines, with particular attention to protocol comparability, execution semantics, and reproducibility. Broader surveys of financial LLM agents~\citep{wu2026agentic,dong2025large} cover applications across finance, while other reviews focus on trading agents, alpha mining, or selected stages of the trading process~\citep{ding2024large,zhang2025survey,okpala2025agentic}.

Despite these advances, three aspects remain insufficiently connected. First, it remains unclear how agents are distributed across the quantitative trading workflow and which stages are less integrated into complete trading systems. Second, existing work provides limited insight into how architecture, coordination, and adaptation connect decisions across trading stages. Third, although prior work has highlighted important differences in evaluation protocols, it remains unclear how evaluation settings relate to different parts of the workflow and what claims about agent capability they can support. We organize the survey around these three questions.

\textbf{Review Scope.} 
To address these questions, we review recent agentic quant trading research, focusing mainly on 2025--2026 while including earlier studies when relevant. We include studies where LLM agents contribute to trading workflows and exclude general financial LLM applications and non-agent prediction models. For complete systems, we code task coverage and system mechanisms using the taxonomies in Sections~\ref{sec:2} and~\ref{sec:3}. The reported counts refer to the studies reviewed here.

\textbf{Our Contributions.} 
Based on this review, we make three main contributions. First, we organize agentic quantitative trading around five stages: factor mining, signal discovery, portfolio construction, order execution, and risk management (Section~\ref{sec:2}). This reveals where agents currently operate across the trading process and where integration remains limited. Second, we analyze complete trading systems through three dimensions: architecture, coordination, and adaptation (Section~\ref{sec:3}), showing how decisions are connected and revised across stages. Third, we compare benchmarks for strategy construction, offline trading, live market evaluation, and reliability assessment (Section~\ref{sec:4}), and derive quantitative findings on workflow coverage, coordination patterns, and evaluation evidence (Section~\ref{sec:5}).

\section{Agentic Workflows across the Quantitative Trading Pipeline}
\label{sec:2}
This section reviews agentic workflows across five stages of the trading pipeline: factor mining, signal discovery, portfolio construction, order execution, and risk management. Factor mining methods are grouped by their dominant agent capability, while the remaining stages are organized by the roles agents play at each stage. Figure~\ref{fig:tree_full} summarizes the resulting framework with representative studies.

\subsection{Agentic Factor Mining}
\label{sec:2.1}

Factor mining aims to discover predictive variables for forecasting asset returns. Existing methods range from manual factor design~\citep{fama2015five} to machine learning~\citep{xu2021hist} and formulaic search \citep{yu2023generating}, but they face limitations such as factor crowding, opacity, redundancy, and weak economic grounding. These challenges motivate agentic factor mining, where LLM-centered agents generate ideas, collaborate through specialized roles, evolve through feedback, and use external tools to execute and evaluate candidate factors.

Recent agentic factor mining studies can be organized around four dominant mechanisms, which are not mutually exclusive. \textbf{(1) Generative agents} transform existing factors, financial rationale, and human ideas into explicit factor formulas. FAMA~\citep{FAMA} emphasizes autonomous exploration. Its Cross-Sample Selection reduces repetitive in-context generation, while Chain-of-Experience reuses successful mining trajectories to guide later search. Alpha-GPT~\citep{Alpha-gpt} instead emphasizes human interaction, allowing quantitative researchers to express investment ideas in natural language and convert them into executable alpha formulas.

\textbf{(2) Collaborative agents} divide factor research among agents with different roles. Instead of assigning ideation, implementation, and validation to a single model, these systems approximate the workflow of a quant research team. FactorMAD \citep{FactorMAD} uses debate between LLM agents to critique and refine alpha factors. AlphaLogics \citep{Alphalogics} organizes collaboration around market-logic discovery, using one set of agents to extract economic rationales and another to generate factors. Across these systems, the value is not merely “more agents,” but the use of structured interaction to improve interpretability, reduce unilateral reasoning errors, and make factor discovery closer to institutional research practice.

\textbf{(3) Self-evolving agents} treat factor mining as an iterative learning process, using memory \citep{FactorMiner, XALPHA}, reinforcement learning (RL) \citep{AlphaAgentEvo}, or trajectory-level evolution \citep{QuantAlpha} to reuse successful patterns and improve future search. The shared logic is that alpha mining in non-stationary markets requires previous experience rather than independent sampling. This makes self-evolving agents promising for long-horizon discovery and redundancy control but also raises evaluation concerns: memory and evolution can overfit to historical validation regimes unless tested under strict out-of-sample and cross-market protocols.

\textbf{(4) Tool-augmented agents} connect LLM reasoning with computational tools that search, implement, and evaluate candidate factors. LLM-MCTS~\citep{LLM-MCTS} combines generation with tree search, AlphaPROBE~\citep{AlphaPROBE} uses graph-guided retrieval, and Hubble~\citep{Hubble} introduces constrained tools to improve the validity and diversity of mined factors. In these systems, tools are not auxiliary components but part of the factor development process. Overall, agentic methods turn factor mining from isolated formula generation into a research workflow that supports collaboration, repeated refinement, and systematic evaluation.

\subsection{Agentic Trading Signal Discovery}
\label{sec:2.2}
Traditional methods typically follow a modular pipeline: predictive factors are designed from prices, fundamentals, or alternative data, then used by statistical and deep learning models to forecast returns \citep{4.0}. Agentic workflows instead allow agents to combine market information across modalities and leverage LLMs' reasoning and generation capabilities to refine trading signals or strategy logic \cite{FinVision,MountainLion}. Since signal discovery is often embedded in e2e agentic trading systems, we focus here on representative signal designs and leave broader system architectures to Section~\ref{sec:3}.

Agents support trading signal discovery in three main ways. \textbf{(1) Agent reasoning signals} treat the LLM agent as the primary producer of trading judgments. Rather than first constructing a formal factor or model, these systems ask agents to absorb heterogeneous market evidence and directly issue recommendations, ratings, or actions. Specialist agents analyze news, fundamentals, price dynamics, and macro conditions, while a synthesis agent converts their outputs into stock-level theses and ordinal recommendations \cite{MarketSenseAI}. Recent studies also introduce simulated investment teams \cite{Tradingagents}, memory layers \cite{Tradinggpt,Finmem}, or verbal reinforcement \cite{yu2024fincon} to stabilize the final buy/sell/hold decision. The strength of this paradigm lies in its ability to combine weak, heterogeneous, and partially textual evidence that is difficult to encode in a fixed model. However, this flexibility also increases the risk of temporal leakage, making backtest results overly optimistic.

\textbf{(2) Factor-model signals} preserve the conventional quantitative premise that signals should come from explicit factors or models while using agents to automate and improve the research process around them. RD-Agent(Q) is representative: it coordinates factor discovery and model innovation, treating signal quality as the result of joint factor-model optimization \citep{RD-Agent-Q}. Compared with direct agent generation, this paradigm offers stronger compatibility with established quantitative practice: signals are more measurable, backtestable, and decomposable. However, its value depends on whether the agent improves the factor/model pipeline rather than merely wrapping it in natural-language explanations.

\textbf{(3) Executable strategy signals} refer to trading signals generated from executable trading logic constructed or optimized by agents. In this paradigm, the agent usually does not directly produce buy/sell/hold decisions through natural-language reasoning. Instead, it first translates trading intentions, market hypotheses, or historical feedback into rules, query languages \cite{OQL}, strategy programs \cite{FactFin}, or trading policies \cite{Flag-trader}. These executable strategy logics are then evaluated in simulated trading environments \cite{QuantAgents} or reinforcement learning environments \cite{alphaquanter, Flag-trader}, and the resulting actions are used as trading signals.

\begin{figure*}[t]
\centering
\includegraphics[width=0.92\textwidth]{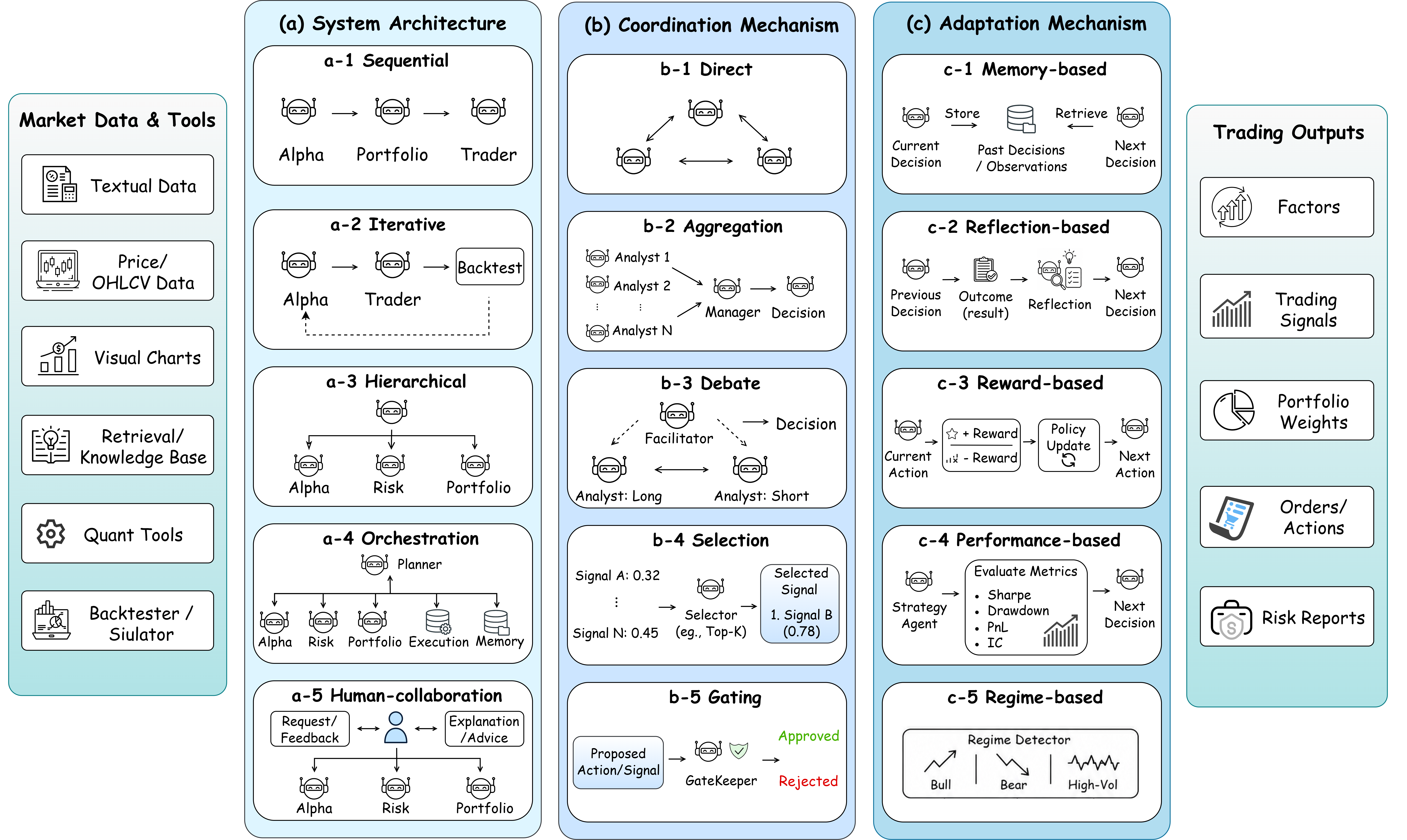}
\caption{Taxonomy of agentic quantitative trading systems across three system-design dimensions.}
\label{fig:e2e_arch}
\end{figure*}

\subsection{Agentic Portfolio Construction}
\label{sec:2.3}
Portfolio construction converts trading signals into capital allocations under risk and diversification constraints. Classical methods such as mean-variance optimization, Black--Litterman allocation~\citep{B_L}, and RL~\citep{jiang2017deep} provide well-established foundations. Recent agentic workflows extend this line of research by introducing specialized agents for asset selection and weight optimization.

\textbf{Portfolio asset selection} concerns which assets should enter the portfolio. Agents combine financial indicators, news, market narratives~\citep{Crypto-Portfolio-MAS}, and simulated investor preferences~\citep{MASS} to produce asset rankings~\citep{3S-Trader} and inclusion decisions. Their main role is to compare heterogeneous evidence rather than solve the numerical allocation problem. This allows textual information and investment rationales to influence portfolio construction. However, the resulting rankings may change with the available evidence and inherit biases from agent judgments. Many systems also rely on simple allocation rules after selecting the assets.

\textbf{Portfolio weight optimization} determines how capital is allocated across selected assets. RAPTOR~\citep{RAPTOR} converts confidence-weighted agent views into inputs for a Black--Litterman optimizer, while CEF-Agents~\citep{CEF-Agents} uses LLM agents to generate heuristic procedures for mean--variance optimization and efficient frontier construction. In both cases, agents provide the views or optimization logic, and conventional optimization methods apply portfolio constraints and produce the final weights.

\subsection{Agentic Order Execution}
\label{sec:2.4}
Traditional order execution converts trading signals and portfolio decisions into market orders, limit orders, rebalancing trades, and execution schedules. Classical methods optimize transaction costs, market impact, slippage, latency, and inventory risk under predefined execution models~\citep{shimoshimizu2024introduction}.

Recent agentic approaches address different parts of this process. \citet{Instruction-to-Order} examine how LLMs interpret incomplete trading instructions, request missing details, and convert them into standardized orders. AlphaCrafter~\citep{AlphaCrafter} translates factor ensembles into portfolio positions and rebalancing orders under risk constraints, while AgenticAITA~\citep{AgenticAITA} uses multi-agent deliberation and deterministic safety checks before execution. PACE~\citep{PACE} extends LLM reasoning to parent-order execution through long-horizon planning and short-horizon adjustment to recent market observations. Together, these studies move agentic trading from deciding what to trade toward controlling how decisions are executed. However, most evidence still comes from backtests, historical replay, or dry runs, with limited testing under live liquidity and execution costs.

\subsection{Agentic Risk Management}
\label{sec:2.5}
Traditional trading systems manage risk through volatility estimation, portfolio constraints, stop-loss rules, and post-trade evaluation~\citep{uryasev2000conditional}. In agentic workflows, risk management has two roles: constraining risky actions before execution and using observed outcomes to adjust future risk decisions.

\textbf{Agentic Risk Control} estimates risk and uses it to constrain trading actions. Trade-or-not~\citep{Trade-or-not} uses agents to identify stochastic models and derive risk measures for trading decisions. WebCryptoAgent~\citep{WebCryptoAgent} separates slower strategic analysis from rapid monitoring during market shocks. At the execution stage, SAE~\citep{SAE} applies exposure budgets, cooldown periods, slippage limits, and venue restrictions before an order is released. Risk information therefore changes the trading action before it reaches the market, rather than appearing only in later performance metrics.

\textbf{Agentic Risk Adaptation} uses past outcomes to revise later risk decisions. FinMem~\citep{Finmem} stores important events and trading outcomes in layered memory, then adjusts its risk stance as market conditions change. ATLAS~\citep{ATLAS} updates the central agent's prompt using delayed and noisy market feedback. TrustTrade~\citep{TrustTrade} uses reflective memory and agreement among agents to reduce the influence of unreliable evidence.

\section{Agentic Quantitative Trading Systems}
\label{sec:3}
While Section~\ref{sec:2} examines how agents support individual stages of the trading pipeline, this section turns to quant trading systems and considers how these stages are connected within agentic workflows. We classify these systems along three dimensions: system architecture, coordination mechanism, and adaptation mechanism, as illustrated in Figure~\ref{fig:e2e_arch}. Table~\ref{tab:e2e} compares representative systems by their coverage of the five pipeline stages and their implementation along the three system dimensions.

\providecommand{\halfc}{\ensuremath{\triangle}}
\providecommand{\na}{\textit{N/A}}
\begin{table*}[t]
\centering
\small
\setlength{\tabcolsep}{2.0pt}
\renewcommand{\arraystretch}{1.1}
\providecommand{\paperentry}[2]{}
\renewcommand{\paperentry}[2]{\parbox[t]{\linewidth}{\raggedright #1\nobreak\hspace{0.15em}\mbox{\citep{#2}}}}
\definecolor{e2erow}{gray}{0.93}
\makebox[\textwidth][l]{%
\begin{tabular}{@{}p{2.2cm} ccccc >{\centering\arraybackslash}p{2.7cm} >{\centering\arraybackslash}p{2.2cm} >{\centering\arraybackslash}p{2.7cm} p{5.7cm}@{\hspace{0.1cm}}}
\toprule
\multirow{2}{2.2cm}{\centering System} &
\multicolumn{5}{c}{Task Coverage} &
\multicolumn{3}{c}{System Technical Implementation} &
\multirow{2}{5.7cm}{\centering Distinctive Mechanism} \\
\cmidrule(lr){2-6}\cmidrule(lr){7-9}
& F & S & P & E & R & Architecture & Coordination & Adaptation & \\
\midrule  
\multicolumn{10}{l}{\textit{\textbf{Panel A: Single-agent systems}}} \\
\rowcolor{e2erow}
\paperentry{FinMem}{Finmem} & -- & \(\checkmark{}\) & -- & \halfc{} & \halfc{} & Sequential, Iterative & \na & Mem., Reflection, Perf. & Layered memory with adaptive risk persona. \\
\paperentry{FinAgent}{FinAgent} & -- & \(\checkmark{}\) & -- & \halfc{} & \halfc{} & Sequential, Iterative & \na & Memory, Reflection & Dual-level reflection with memory retrieval. \\
\rowcolor{e2erow}
\paperentry{FLAG-Trader}{Flag-trader} & -- & \(\checkmark{}\) & -- & \halfc{} & -- & Iterative & \na & Reward & PPO-tuned LLM policy network. \\
\midrule
\multicolumn{10}{l}{\textit{\textbf{Panel B: Multi-agent systems}}} \\
\rowcolor{e2erow}
\paperentry{TradingGPT}{Tradinggpt} & -- & \(\checkmark{}\) & \halfc{} & \halfc{} & \halfc{} & Iterative & Agg., Debate & Memory, Reflection & Layered memory agents with peer debate. \\
\paperentry{FinCon}{yu2024fincon} & -- & \(\checkmark{}\) & \(\checkmark{}\) & \halfc{} & \(\checkmark{}\) & Iterative, Hierarchical & Agg. & Mem., Reflection, Perf. & Dual-level risk control with CVRF belief updates. \\
\rowcolor{e2erow}
\paperentry{FinVision}{FinVision} & -- & \(\checkmark{}\) & \halfc{} & \halfc{} & \halfc{} & Sequential, Iterative & Aggregation & Reflection, Perf. & Visual reflection with portfolio-percent sizing. \\
\paperentry{ContestTrade}{ContestTrade} & \(\checkmark{}\) & \(\checkmark{}\) & \(\checkmark{}\) & \halfc{} & \halfc{} & Sequential, Hierarchical & Agg., Select & Performance & Market-feedback contests for factor selection. \\
\rowcolor{e2erow}
\paperentry{Orchestration}{li2025orchestration} & \(\checkmark{}\) & \(\checkmark{}\) & \(\checkmark{}\) & \(\checkmark{}\) & \(\checkmark{}\) & Hierarchical, Orch. & Direct, Agg., Gate & Mem., Perf., Regime & MCP/A2A from alpha to execution. \\
\paperentry{FinArena}{Finarena} & -- & \(\checkmark{}\) & \halfc{} & \halfc{} & \halfc{} & Hierarchical, Human & Aggregation & Reflection & Human risk preferences in MoE-style forecasts. \\
\rowcolor{e2erow}
\paperentry{MountainLion}{MountainLion} & -- & \(\checkmark{}\) & \(\checkmark{}\) & \halfc{} & \halfc{} & Orch., Human & Agg. & Refl., Perf., Regime & GraphRAG components for horizon-aware advice. \\
\paperentry{QuantAgent-P}{QuantAgent-Price} & -- & \(\checkmark{}\) & -- & \halfc{} & \(\checkmark{}\) & Sequential, Orch. & Aggregation & Regime & Price-driven HFT agents with risk bounded orders. \\
\rowcolor{e2erow}
\paperentry{QuantAgents}{QuantAgents} & -- & \(\checkmark{}\) & \(\checkmark{}\) & \halfc{} & \(\checkmark{}\) & Iterative, Hierarchical & Agg., Debate & Memory, Performance, Reflection, Reward & Simulated trading meetings with dual rewards. \\
\paperentry{QuantEvolve}{QuantEvolve} & \halfc{} & \(\checkmark{}\) & \halfc{} & \halfc{} & \halfc{} &  Iterative, Orch. & Agg., Select & Performance, Regime & Quality-diversity evolutionary strategy archive. \\
\rowcolor{e2erow}
\paperentry{RD-Agent(Q)}{RD-Agent-Q} & \(\checkmark{}\) & \(\checkmark{}\) & -- & -- & -- & Iterative, Orch. & Direct, Select & Mem., Perf., Reward & Joint factor–model optimization. \\
\paperentry{TradingAgents}{Tradingagents} & -- & \(\checkmark{}\) & -- & \halfc{} & \(\checkmark{}\) & Sequential, Hierarchical & Agg., Debate & Reflection & Trading firm roles with debate and approval. \\
\rowcolor{e2erow}
\paperentry{TradingGroup}{TradingGroup} & -- & \(\checkmark{}\) & \halfc{} & \halfc{} & \(\checkmark{}\) & Sequential, Iterative & Aggregation & Mem., Reflection, Perf. & End-to-end data-synthesis pipeline. \\
\paperentry{AgenticAITA}{AgenticAITA} & -- & \(\checkmark{}\) & \halfc{} & \(\checkmark{}\) & \(\checkmark{}\) & Sequential, Orch. & Direct, Gate & Memory, Regime & Typed JSON contracts, deterministic hard gates. \\
\rowcolor{e2erow}
\paperentry{AlphaCrafter}{AlphaCrafter} & \(\checkmark{}\) & \(\checkmark{}\) & \(\checkmark{}\) & \(\checkmark{}\) & \(\checkmark{}\) & Sequential, Iterative & Agg., Select & Mem., Perf., Regime & Continuously adaptive factor-to-execution. \\
\paperentry{Invest-Teams}{Expert-Investment-Teams} & -- & \(\checkmark{}\) & \(\checkmark{}\) & -- & \halfc{} & Hierarchical & Agg., Select & \na & Explicitly decomposes into fine-grained tasks. \\
\rowcolor{e2erow}
\paperentry{HedgeAgents}{HedgeAgents}
& -- & \(\checkmark{}\) & \(\checkmark{}\) & \halfc{} & \(\checkmark{}\)
& Iterative, Hierarchical
& Direct, Agg.
& Mem., Refl., Perf.
& Hedging experts coordinated through conferences. \\
\bottomrule
\end{tabular}%
}
\caption{Comparison of representative agentic quantitative trading systems across pipeline coverage and system mechanisms. F/S/P/E/R denote the five trading-workflow stages defined in Section 2. \(\checkmark{}\) denotes core coverage, \halfc{} partial coverage, -- no coverage; N/A indicates that no corresponding mechanism was identified.}
\label{tab:e2e}
\end{table*}

\subsection{System Architecture}
\label{sec:3.1}
\textit{How are agentic workflows organized?}
Agentic trading systems differ in both what individual agents do and how they are arranged within the workflow. System architecture defines the order in which agents act and who controls the workflow. These choices determine how market analysis, portfolio construction, and execution are linked.

\textbf{Sequential} and \textbf{iterative} structures are the most common and are frequently combined. Sequential pipelines organize trading tasks in a fixed order, which makes it easier to inspect the output of each stage and trace how a decision was produced. However, problems found at later stages may not prompt earlier agents to revise their decisions. Iterative architectures address this limitation by returning execution feedback to earlier stages, often through shared memory. Agents can then revise trading plans as market conditions change. QuantEvolve~\citep{QuantEvolve} iterates through hypothesis generation, coding, backtesting, and archive updates to retain strategies that perform well while maintaining diversity. RD-Agent(Q)~\citep{RD-Agent-Q} alternates between research and development, using backtest feedback to jointly refine factors and models while a multi-armed bandit scheduler selects the next direction.

\textbf{Hierarchical} and \textbf{orchestration} architectures are closely related because both introduce a component that coordinates specialist agents. In a hierarchical system, a manager assigns tasks, reviews the agents' outputs, and makes or approves the final decision. QuantAgents~\citep{QuantAgents}, for example, uses a manager and analyst structure to combine specialist analysis through repeated deliberation. Orchestration extends this role from decision making to workflow control. The orchestrator determines which agent or tool runs next, passes results between modules, and repeats a step when its output fails a check. The system proposed by~\citet{li2025orchestration} uses MCP and A2A protocols to connect research, portfolio construction, risk control, and execution. In simple terms, a hierarchical manager decides among agent recommendations, whereas an orchestrator controls how the workflow proceeds. A system may use both structures. 

\textbf{Human collaboration} can be added to any architecture by reserving selected inputs or decisions for users. In FinArena~\citep{Finarena}, investors specify their risk preferences through an interactive interface, and the agent system uses them to produce personalized investment decisions.

\subsection{Coordination Mechanism}
\label{sec:3.2}
\textit{How do LLM agents make decisions?}
In agentic quantitative trading systems, coordination mechanisms determine how their outputs are combined into a trading decision. They govern how information passes between agents, how disagreements are handled, and what checks an action must pass before execution.

\textbf{Direct} coordination and \textbf{aggregation} govern how information flows between agents. Direct coordination sends one agent's output to the next agent or module without first combining it with other results. In the framework by Li et al.~\citep{li2025orchestration}, A2A enables agents to exchange messages directly, while MCP carries instructions between the orchestrator and individual agents. Aggregation instead collects outputs from several agents and combines them into one decision. It is widely used in systems such as ContestTrade, TradingGroup, and AlphaCrafter, where specialist outputs are combined before the next trading stage~\citep{ContestTrade,TradingGroup,AlphaCrafter}. Direct coordination preserves the original output and allows fast transfer, while aggregation brings together several views but may lose detail when they are summarized.

\textbf{Debate} and \textbf{selection} handle disagreement in different ways. Debate allows agents to challenge each other's evidence before a decision is made. QuantAgents~\citep{QuantAgents} organizes trading meetings in which specialist agents present different views and a manager makes the decision. Because each agent and discussion round requires more model calls, debate is better suited to major portfolio decisions than rapid execution. Selection instead compares candidate outputs and retains those with stronger evaluation results. ContestTrade~\citep{ContestTrade} ranks agents using market feedback, giving greater influence to agents with better past performance. AlphaCrafter~\citep{AlphaCrafter} uses validation results to select agents for the final ensemble. Selection requires less interaction than debate, but repeated ranking may favor the same agents and reduce strategy diversity.

\textbf{Gating} checks whether an agent output meets required rules before it proceeds to the next stage. In AgenticAITA~\citep{AgenticAITA}, typed JSON contracts specify what each stage must return, while a deterministic hard gate blocks outputs that violate safety rules before execution. This makes rejected actions easier to identify and trace, although strict gates may also block useful outputs that do not match predefined requirements.
\begin{table*}[t]
\centering
\footnotesize
\setlength{\tabcolsep}{1.45pt}
\renewcommand{\arraystretch}{1}

\providecommand{\paperentry}[2]{#1~\citep{#2}}
\providecommand{\tableentry}[1]{#1}
\providecommand{\headentry}[1]{\centering\textbf{#1}\arraybackslash}

\definecolor{benchmarkrow}{gray}{0.93}

\begin{tabular}{
>{\raggedright\arraybackslash}m{0.11\textwidth}
>{\raggedright\arraybackslash}m{0.15\textwidth}
>{\raggedright\arraybackslash}m{0.15\textwidth}
>{\raggedright\arraybackslash}m{0.15\textwidth}
>{\raggedright\arraybackslash}m{0.41\textwidth}
}
\toprule
\headentry{Paper} &
\headentry{Task} &
\headentry{Data} &
\headentry{Metrics} &
\headentry{Main Findings} \\
\midrule

\multicolumn{5}{l}{\textit{\textbf{Panel A: Strategy Construction Evaluation}}} \\

\rowcolor{benchmarkrow}
\paperentry{AlphaEval}{AlphaEval} &
\tableentry{Formulaic alpha evaluation} &
\tableentry{A-shares, US stocks} &
\tableentry{Predictiveness, stability, robustness, diversity} &
\tableentry{Multidimensional metrics identify high quality alphas more effectively than backtest-only screening.} \\

\paperentry{QuantCode}{QuantCodeBench} &
\tableentry{Strategy code generation} &
\tableentry{New task-specific datasets} &
\tableentry{Judge pass rate} &
\tableentry{Reliable strategy generation requires runnable code, realistic trading logic, and alignment with user intent.} \\

\rowcolor{benchmarkrow}
\paperentry{OQL}{OQL} &
\tableentry{Natural language to option strategies} &
\tableentry{New task-specific datasets} &
\tableentry{Query quality \& strategy quality} &
\tableentry{Semantic accuracy predicts strategy quality better than execution success alone, while OQL reduces hallucination and risk.} \\

\midrule
\multicolumn{5}{l}{\textit{\textbf{Panel B: Offline Trading Evaluation}}} \\

\rowcolor{benchmarkrow}
\paperentry{InvestorBench}{InvestorBench} &
\tableentry{Financial decision-making} &
\tableentry{Stocks, crypto, ETFs; prices, reports, news} &
\tableentry{CR, MDD, Sharpe, volatility} &
\tableentry{Agents that incorporate advanced memory systems and dynamic risk assessment improve adaptation to market fluctuations.} \\

\paperentry{StockAgent}{StockAgent} &
\tableentry{Stock market simulation} &
\tableentry{US stocks; reports, BBS, events} &
\tableentry{Effectiveness, reliability, impact} &
\tableentry{External information, agent profiles, and backbone models shape trading behavior and simulated price dynamics.} \\

\rowcolor{benchmarkrow}
\paperentry{StockBench}{StockBench} &
\tableentry{Contamination-free stock back-trading} &
\tableentry{US stocks; prices, fundamentals, news} &
\tableentry{Return, MDD, Sortino} &
\tableentry{Most agents struggle to beat buy and hold, and strong static financial knowledge does not guarantee trading success.} \\

\midrule
\multicolumn{5}{l}{\textit{\textbf{Panel C: Live Market Evaluation}}} \\

\rowcolor{benchmarkrow}
\paperentry{DeepFund}{DeepFund} &
\tableentry{Live fund investment} &
\tableentry{US stocks; prices, reports, fundamentals, news} &
\tableentry{CR, CRbnh, MDD, Sharpe, WR, beta, alpha} &
\tableentry{Most LLMs lost money in live evaluation; cash management and diversification were more decisive than fluent analysis.} \\

\paperentry{AI-Trader}{AITrader} &
\tableentry{Autonomous live trading} &
\tableentry{A-shares, US stocks, crypto} &
\tableentry{CR, MDD, Sortino, volatility} &
\tableentry{General intelligence does not transfer reliably to trading; risk control drives cross-market robustness.} \\

\rowcolor{benchmarkrow}
\paperentry{LiveTradeBench}{LiveTradeBench} &
\tableentry{Live portfolio allocation} &
\tableentry{US stocks, Polymarket} &
\tableentry{CR, WR, MDD, Sharpe, volatility} &
\tableentry{Models show distinct allocation styles; strong static benchmark scores do not predict live trading performance.} \\

\paperentry{AMA}{AgentMarketArena} &
\tableentry{Lifelong live trading arena} &
\tableentry{US Stocks, crypto; prices, news} &
\tableentry{CR, MDD, Sharpe, volatility} &
\tableentry{Agent architecture explains more performance variation than the LLM backbone under shared verified inputs.} \\

\rowcolor{benchmarkrow}
\paperentry{PolyBench}{PolyBench} &
\tableentry{Prediction-market forecasting and trading} &
\tableentry{Polymarket; news} &
\tableentry{Return, Sharpe, accuracy} &
\tableentry{Accuracy and confidence do not ensure profit; order-book liquidity and slippage materially change rankings.} \\

\paperentry{FinDF}{FinDeepForecast} &
\tableentry{Live financial forecasting} &
\tableentry{Economies, companies; macro tasks} &
\tableentry{Accuracy} &
\tableentry{Deep research agents lead, but precise recurrent numerical forecasting remains difficult under temporal isolation.} \\

\midrule
\multicolumn{5}{l}{\textit{\textbf{Panel D: Reliability Evaluation}}} \\

\rowcolor{benchmarkrow}
\paperentry{FinLake-Bench}{FactFin} &
\tableentry{Leakage and counterfactual evaluation} &
\tableentry{Six assets; prices, news, counterfactuals} &
\tableentry{CR, MDD, Sharpe, PC, confidence invariance} &
\tableentry{Backtest gains fall after model cutoffs, while counterfactual tests reveal reliance on memorized outcomes.} \\

\paperentry{AutoRedTrader}{AutoRedTrader} &
\tableentry{Adversarial misinformation robustness} &
\tableentry{BTC; prices, news, synthetic misinformation} &
\tableentry{CR, exposure \& success rate} &
\tableentry{Subtle financial misinformation can change trading decisions, while time-series grounding reduces attack effectiveness.} \\

\rowcolor{benchmarkrow}
\paperentry{KTD-FIN}{KTD_FIN} &
\tableentry{Leakage-controlled attribution benchmark} &
\tableentry{A-shares} &
\tableentry{CR, MDD, Sharpe, turnover, ECE} &
\tableentry{Masking alters agent decisions, while returns mainly reflect market and style exposure rather than stock selection.} \\

\bottomrule
\end{tabular}

\caption{Benchmark and dataset comparison for agentic quantitative trading systems. Abbreviations: CR = cumulative return; WR = win rate; MDD = maximum drawdown; PC = prediction consistency, ECE = expected calibration error}
\label{tab:db}
\vspace{-4.5mm}
\end{table*}

\subsection{Adaptation Mechanism}
\label{sec:3.3}
\textit{How do LLM agents change over time?}
Financial markets change over time, and profitable signals often weaken. Trading agents therefore need mechanisms that revise their behavior as new evidence becomes available.

\textbf{Memory} stores earlier market evidence and trading outcomes for later retrieval. FinMem~\citep{Finmem} organizes financial information through layered memory and adjusts its memory span to retain important evidence. \textbf{Reflection} instead reviews past outcomes and extracts lessons for future decisions. FinAgent~\citep{FinAgent} links market evidence with subsequent price movements, then evaluates earlier trades and recommends corrections. TradingGroup~\citep{TradingGroup} similarly allows its forecasting, style, and decision agents to learn from past successes and failures. These mechanisms make changes easier to trace, although noisy returns may still obscure which decision caused a gain or loss.

 \textbf{Performance} mechanisms return trading results to the agents. In FinVision~\citep{FinVision}, a reward agent calculates performance metrics that support the next round of prediction and reflection. \textbf{Reward} mechanisms convert outcomes into signals that shape future behavior. FinCon~\citep{yu2024fincon} extracts investment beliefs from changes in profit and loss, then sends verbal feedback to the agents that require revision. \textbf{Regime }mechanisms adjust the strategy according to the detected market state. AlphaCrafter~\citep{AlphaCrafter} uses a Screener agent to assess the current regime and construct a matching factor ensemble. Performance feedback is most useful when interpreted together with regime signals, because weaker results may reflect a change in market style rather than a decline in predictive ability.

\section{Datasets and Benchmarks}
\label{sec:4}

Datasets and benchmarks determine which parts of an agentic trading workflow can be evaluated and how reliable the reported results are. Table~\ref{tab:db} organizes existing studies into four groups: strategy construction, offline trading, live market evaluation, and reliability evaluation. Strategy construction benchmarks assess the quality of alpha formulas and whether trading intent can be translated into executable code or structured strategies~\citep{AlphaEval,QuantCodeBench,OQL}. Offline benchmarks instead test trading decisions with historical data or simulated markets, allowing controlled comparison across models and agent designs~\citep{InvestorBench,StockAgent,StockBench}.

Recent benchmarks increasingly move from isolated outputs to repeated decisions under live market conditions. DeepFund shows the practical difficulty of generating stable returns in live fund management and highlights the value of cash management and diversification~\citep{DeepFund}. Agent Market Arena further finds that, under shared verified inputs, agent architecture explains more performance variation than the LLM backbone~\citep{AgentMarketArena}. PolyBench shows that high forecasting accuracy and confidence do not guarantee profit once liquidity and slippage affect execution~\citep{PolyBench}. Live evaluation therefore needs to assess trading decisions under actual market conditions rather than language capability alone.

Benchmark design is also shifting from measuring returns to testing whether those returns are trustworthy. Profit Mirage introduces FinLake-Bench to examine whether apparent performance depends on knowledge leakage and memorized market outcomes~\citep{FactFin}. AutoRedTrader tests how synthetic misinformation changes agent decisions and whether time series evidence improves resistance~\citep{AutoRedTrader}. KTD-FIN controls ticker and date memory channels, then separates market exposure, style exposure, and stock selection effects~\citep{KTD_FIN}. These studies show that cumulative return alone cannot establish agent capability. Reported gains are more credible when they survive temporal controls and adversarial tests, with attribution confirming where the returns originate.

\section{Discussion}
\label{sec:5}

The comparison in Tables~\ref{tab:e2e} and~\ref{tab:db} reveals three clear patterns. These counts describe the systems and benchmarks included in this review rather than the entire literature.

\textbf{Finding 1: Signal discovery dominates, while complete trading workflows remain rare.}
All 20 systems in Table~\ref{tab:e2e} treat signal discovery as a core function. Factor mining is core in 4 systems, portfolio construction in 8, order execution in 3, and risk management in 9. Only 4 systems cover all five stages at least partially, and only 2 treat all five as core functions. Most systems can therefore produce trading signals, but few carry them through portfolio construction, execution, and risk control.

\textbf{Finding 2: Most systems mix architectures, but aggregation dominates coordination.}
Seventeen of the 20 systems combine more than one architectural structure, and 12 use an iterative structure. Among the 17 multi-agent systems, 15 aggregate outputs from several agents. Selection appears in 5 systems, debate in 3, and explicit gating in 2. Agent roles and workflow structures have become diverse, but most systems still rely on aggregation to reach a final decision.

\textbf{Finding 3: Evaluation setting changes what the reported results mean.}
The 15 benchmarks in Table~\ref{tab:db} include 3 for strategy construction, 3 for offline trading, 6 for live evaluation, and 3 for reliability. All live-market benchmarks that report trading outcomes show that realized performance depends on factors beyond model capability, including risk control, system design, liquidity, and slippage~\citep{DeepFund,AITrader,LiveTradeBench,AgentMarketArena,PolyBench}. All three reliability benchmarks show that leakage, misinformation, or return attribution can change the interpretation of agent performance~\citep{FactFin,AutoRedTrader,KTD_FIN}. Backtest success or forecasting accuracy therefore provides weaker evidence than live trading or reliability evaluation.

\section{Future Directions}
\label{sec:6}

\textbf{Complete trading workflows.}
Current systems are stronger at generating signals than turning them into portfolios, orders, and risk decisions. Future work should evaluate the full path from signal generation to executed trades. Agents do not need to replace all components; portfolio optimizers, execution algorithms, and risk rules can remain conventional, while agents handle tasks requiring reasoning or changing information~\citep{RAPTOR,AlphaCrafter,li2025orchestration}. Feedback from later outcomes should also update earlier decisions.

\textbf{Coordination beyond aggregation.}
Most multi-agent systems combine outputs, but few test which agent should be trusted or when an action should be blocked. Selection, debate, and gating provide alternative mechanisms. ContestTrade selects agents using past performance, QuantAgents enables competing views, and AgenticAITA blocks unsafe actions before execution~\citep{ContestTrade,QuantAgents,AgenticAITA}. Future studies should compare these mechanisms under the same settings and record why decisions are changed or rejected.

\textbf{Evaluation matched to capability claims.}
Different evaluations support different conclusions. An executable strategy does not guarantee profit, and a profitable backtest does not prove live trading ability. Future benchmarks should match evaluation settings to the capability being tested, while controlling information timing, execution costs, liquidity, and slippage. Profit Mirage, KTD-FIN, and PolyBench show that conclusions can change when leakage, attribution, or market frictions are considered~\citep{FactFin,KTD_FIN,PolyBench}.

\section{Conclusion}
\label{sec:7}

This survey examines agentic quantitative trading from trading tasks to complete systems and their evaluation. The reviewed systems show a clear imbalance: signal discovery is nearly universal, while complete integration with portfolio construction, execution, and risk control remains rare. System architectures are often mixed, but multi-agent decisions still rely heavily on aggregation. Benchmark evidence also becomes less favorable once evaluation moves from historical tests to live markets and reliability controls. These results suggest that the next challenge is not simply to build larger groups of trading agents, but to connect decisions to execution and test them under conditions that reveal whether the resulting returns are real and reproducible.

\bibliographystyle{ACM-Reference-Format}
\bibliography{reference}

\end{document}